\documentstyle [a4,12pt]{article}

\begin{document}
\begin{center}
{\large \bf EXPERIMENTAL CONSEQUENCES
OF THE HYPOTHESIS ABOUT A
FUNDAMENTAL MASS IN HIGH - ENERGY PHYSICS}\\
 \vspace{0,5cm}
 {\bf Rustam M.Ibadov}\\
 \vspace{0,5cm}
 Department of Theoretical Physics and
Computer Sciences, of Samarkand State University, Uzbekistan. 15,
University blvd, 703004 Samarkand, Uzbekistan.\\
ibrustam@samdu.uz; ibrustam@mail.ru
\end{center}
\vspace{0.8cm}

{\bf Abstract}\\
This work continues earlier investigations towards constructing a
consistent new Quantum Field Theory with fundamental mass $M$,
defining a hypothetical but universal scale in the region of
ultrahigh energies. From a theoretical point of view the fundamental
mass $M$ and corresponding to it the fundamental length $\ell
=\hbar/Mc$ are supposed to play a major role such as Planck's
constant $\hbar$, the speed of light $c$ or Newton's gravitational
constant $ \kappa $.

The standard Quantum Field Theory is recovered in the flat limit.
Furthermore, on the basis of this theory various cross sections of
fundamental processes have been calculated. Some results revealed
that the novel interaction induced via the geometric structure of
the momentum space do not keep spirally. As a characteristic feature
this interaction inherently leads to a violations of fundamental
symmetries, such as $P$ and $CP$. Also the
helicity will no longer be preserved at ultra-high energies.\\


{\bf 1. Introduction}\\
 In the present work some experimental
predictions of the Quantum Field Theory (QFT) with fundamental mass
(FM) $M$ (respectively for the fundamental length $\ell =\hbar/Mc$ )
will be revealed. This fundamental mass $M$ and corresponding to it
the fundamental length $\ell =\hbar/Mc$ may play a similar role as
Planck's constant $\hbar$, the speed of light $c$ or Newtonian
gravitational constant $ \kappa $, defining a characteristic scale
in the region of ultrahigh energies.

The existence of so-called ultra-violet divergences, i.e.,
infinitely large values, arising as a result of direct application
of equations QFT in area of very small space-time distances, or
equivalently, to the region of very high energies and pulses is one
of lacks of standard QFT. There were ideas about presence of a new
universal constant dimension of mass or length in nature\cite{1},
which would fix the certain scale in the field of high  energies or
on small space-time distances because of the purpose to give the
decision of this problem in the most various contexts. They testify
only that the modern high-energy physics still far will defend from
that boundary behind which can will be shown new geometrical
properties of space - time.

 From a position of today it is represented to many theorists rather
probable, that the ``true`` theory of the field, capable to give the
adequate description of all interactions of elementary particles,
will be at least renormalized Lagrange theory having local gauge
supersymmetry. It is asked, whether such scheme can contain
parameter such as fundamental length? The future experiments can
give the answer to this question only. However numerous attempts to
construct more general QFT, proceeding from such parcels, did not
give essential results. This failure is probable speaks that for
today the mathematical theory of spaces which geometry only ``in
small`` differs from (pseudo)Euclidean geometry is not developed
almost and, especially, in similar kinds of spaces the mathematical
device adequate to requirements QFT is not advanced. But, the output
from the created situation is prompted by QFT. As it is known,
within the framework of this theory space-time description is
completely equal in rights with the description in terms pulse-power
variables. If the theory is formulated in pulse representation
fields, sources, Green's functions and other attributes of the
theory appear certain in four-dimensional (pseudo)  Euclidean
p-space.  This modified quantum field theory has an elegant
geometrical basis: in momentum representation one faces a momentum
space corresponding to a de Sitter space of curvature radius
\cite{2,3,4,5,6,7,10,13,14,15,16,17,18}. The approach developed
earlier has been based on the assumption that the momentum space
possesses the geometric structure of a de Sitter space of constant
curvature. A key role has been be assigned to this constant radius
of curvature.

 We shall investigate the phenomenological (experimental) consequences
 of such a quantum field
theoretical model. Calculations of cross sections corresponding to
various basic processes will be carried out up to the second order.
On the basis of the QFT with FM \cite{3,4,5,6,7,10,13,16,17},
calculations of cross sections for processes  such as $ e^{-}e^{-}
\to e^{-}e^{-}$ , $ e^{-}e^{+} \to e^{-}e^{+}$ and $ e^{-}e^{+} \to
\mu^{-} \mu^{+} $ have been carried out by taking into account the
polarization of particles. The contribution of the FM could be
revealed. Some experimental consequences \cite{4, 5, 13, 16,17} are
predicted. In all these calculations the polarization of particles
will be taken into account.

 Experimental
detection of the new fundamental scale testifying to existence
specific atoms of space - time, would mean, that in knowledge of a
nature the new step, commensurable on the value with opening quantum
properties of a matter is made. According to modern data, if the
constant  $ \ell$ also exists then if submits to
restriction\footnote {For the investigation of the inner structure
of hybrid stars in Ref. \cite{19} it is also introduced the fifth
dimension. Because of the extra motion in the fifth dimension, an
extra mass term appears in $4D$ description. The length of $\sim
10^{-12} - 10^{-13} cm$, implied He extra ``mass``  $ m \sim 100
Mev$}  $ \ell\leq 10^{-19} $ cm. This boundary still extremely far
will defend from `` Planck lengths `` $ \ell_{Planck}\sim 10^{-33} $
cm, determining spatial scales of effects of quantum gravitation.
And, certainly, it is impossible to exclude, that in process of
overcoming an enormous interval $ 10^{-19}-10^{-33} $ cm will be
open the new physical phenomena and laws, associate with new ``
scale of a nature `` - fundamental length $ \ell$ .

{\bf 2. ``Fundamental`` equation with universal scales}\\
 In the papers cited \cite{2,3,6,7}, the key role was played
by the following geometric idea: to construct QFT providing an
adequate description of particle interactions at super high
energies, one should write down the standard field theory in the
momentum representation, and then pass it from the Minkowski p-space
to the de Sitter p-space with a large enough radius.

 The de Sitter space has a constant curvature. Depending on its sign there
are two possibilities
\begin{equation} \begin{array}{ccc}
p^{2}_{0}- p^{2}_{1}- p^{2}_{2}- p^{2}_{3}+ p^{2}_{5}\equiv g^{KL}P_{K}P_{L}=M^{2};\\
\hspace{1cm}\small {K,L=0,1,2,3,5}\\
(positive\  curvature: \quad
g^{00}=-g^{11}=-g^{22}=-g^{33}=g^{55}=1) \end{array}
\end{equation}
\begin{equation}
\begin{array}{ccc}
p^{2}_{0}- p^{2}_{1}- p^{2}_{2}- p^{2}_{3}- p^{2}_{5} \equiv g^{KL}P_{K}P_{L}=-M^{2};\\
\hspace{2cm}\small {K,L=0,1,2,3,5}\\
(negative\  curvature: \quad
g^{00}=-g^{11}=-g^{22}=-g^{33}=-g^{55}=1).
\end{array}
\end{equation}

Unlike the previous papers of this sequel, the new theory is being
constructed from the beginning in the configuration representation
\cite{10}.

 The non-Euclidean Lobachevsky imaginary 4-space (2) is
also called the Lobachevsky imaginary 4-space \cite{11}. It is
natural that QFT based on momentum representation of the form
(l)-(2) must predict new physical phenomena at energies $E\ge $M. In
principle, the parameter may turn out to be close to the Planck mass
$M_{Planck}=\sqrt{{\hbar c\over \kappa }}\approx 10^{19}$ GeV. Then,
the new scheme should include quantum gravity. The standard QFT
corresponds to the ``small`` 4-momentum approximation
$$
\left|p_{0}\right| ,\left|\vec{p}\right|\ll  M \quad
p^{5}=g^{55}p_{5}\cong  M ,
$$
which formally can be performed by letting $M\rightarrow \infty $
(``flat limit``). Such features of the considered generalization of
the theory as geometricity and minimality are intriguing. This is
due to the fact that the Minkowski momentum 4-space having a
constant zero curvature is a degenerate limiting form of each of the
spaces with constant nonzero curvature (l)-(2). The formulation of
QFT with fundamental mass discussed in this paper is based on the
quantum version of the de Sitter equation (2) i.e. on the
five-dimensional field equation
\begin{equation}
\left[\frac{\partial^2}{\partial x^{\mu}\partial x_{\mu}}-
\frac{\partial^2}{\partial x^2_5}-\frac{M^2c^2}{\hbar^2}\right]
\Phi(x,x^5)=0,   \small{\qquad \mu =0,1,2,3.}
\end{equation}
derived from (2, using the following substitution, standard for
quantum theory:
$$ p_{\mu }=i\hbar {\partial \over \partial x^{\mu }}$$ and
$$p_{5}=i\hbar {\partial \over \partial x^{5}}.
$$
We deliberately use in (3)the normal units to emphasize those three
universal constants $\hbar$, $c$ and $M$ are grouped into one
parameter - fundamental length $\ell =\hbar/Mc$ . Eq. (3) may be
considered as the ``fundamental`` equation of motion. It is natural
to extend the term ``fundamental`` to Eq. (3) itself (for short
f.e.). All the fields independent of their tensor (or spinor)
character must obey Eq. (3) since similar universality is inherent
in the ``classical`` prototype, i.e. - De Sitter p-space (2). As
applied to scalar, spinor, vector and other fields we shall write
down the five-dimensional wave function $\Phi(x,x^5) $ in the form
$\varphi (x,x^{5} ), $ $\psi _{\alpha } (x,x^{5} )$ and $A_{\mu }
=(x,x^{5} )$ . The field theory based on f.e. (3) turns out to be
more consistent and more general than the scheme developed in the de
Sitter p-space (2). Thus, by virtue of (2) the 4-momentum components
should obey the constraint
$$p^{2}= p^{2}_{0}- \vec{p}^{ \hspace{0.2cm}{2}}\ge  -M^{2} ,
$$
which does not follow from Eq. (3). We should like to note that
having placed the Cauchy problem E.q. (3) for with respect to the
coordinate $x^5$ and just in the correct formulation in the basis of
QFT, in fact, we have introduced a new concept of the field, which
is not equivalent to the notion of the field developed in the theory
with constant curvature momentum space. Consequently, there is a
one-to-one correspondence
$$\Phi(x,x^5)\leftrightarrow \left(
\begin{array}{c}
\Phi(x,0) \\
\partial \Phi(x,0)/\partial x^{5} .
\end{array}
\right)
$$ In other words, the statement that to each field in the 5-space
there corresponds its wave function $\Phi(x,x^5)$ , obeying E.q.
(3), implies that each of these fields in the usual space-time is
described by the wave function with a doubled number of components
$$\left(
\begin{array}{c}
\Phi(x,0) \\
\partial \Phi(x,0)/\partial x^{5}\end{array}\right)=\left(
\begin{array}{c} \Phi_{1}(x)\\ \Phi_{2}(x)
\end{array}
\right).
$$
From here we see that fields are doubled. It appears the field
$\Phi_{2}(x)$ participates only in interactions. Due to it there are
new members in diagrams. Then, it is natural to assume that the
initial data obey the Lagrangian equations of motion following from
the action principle
$$
S=\int d^{4}x L\left[ \Phi (x,0),{\partial \Phi (x,0)\over \partial
x^{5}}\right].
$$
The basic problem of the new theory was is to construct explicit
expressions for the Lagrangians
$$
L\left[\Phi (x,0),{\partial \Phi (x,0)\over \partial x^{5}}\right]
$$
in physically interesting cases, to clarify the meaning of
additional field variables and to extract new physical effects in
the region of super-high energies $E \geq M$ . Partially, this
problem has been solved earlier \cite{2,6,7}. Thus, a doubled number
of field degrees of freedom specific of the new scheme disappears as
M$ \rightarrow \infty $  . Hence, specifically,
$$ \lim_{M \rightarrow \infty } L\left [\Phi (x,0),
{\partial \Phi (x,0)\over \partial x^{5}}\right]=L\left[\Phi
(x,0)\right].
$$
Certainly, if in formulating the Cauchy problem we imposed the
initial conditions at an arbitrary fixed value $ x^5=const$, then
all our conclusions would be the previous ones and the formulae
would undergo trivial changes. For instance, there would appear the
following expression for the action
$$
S=\int_{x^5=const} d^{4}x L\left[\Phi (x,x^5), {\partial \Phi
(x,x^5)\over \partial x^{5}}\right] .
$$
Thus, in Ref. \cite{10} we receive actions for scalar, Dirac and
vector fields. Basically symmetry of the equation of motion -
simultaneously symmetry of action. Therefore it is satisfied
$$\partial S/ \partial x^{5} =0.
$$

{\bf 3. Results and Conclusion}\\
Could the advanced theory be free from ultra-violet divergences? At
the present we do not have the final answer what this issue is
concerned, however, we can calculate effective cross sections of
some processes which are in good agreement with experiments, and
this allows to estimate the contribution of fundamental mass. We
investigate various aspects QFT with FM. It is shown that the
contribution of the auxiliary fields to the Lagrangians of the QFT
with FM is such that the sum of the kinetic terms corresponding to
scalar and Majorana fields is invariant under suppersymmetry
transformations inherent in the Wess-Zumino model \cite{8, 9}. It is
offered on basis QFT with FM the Rotation Invariant Gauge Model with
the compact momentum space \cite{12,18}. Concept of Stochastic
Quantization of the Abelian Fields and Fundamental Mass \cite{15}.
But, the main attention is given to application of our theory to
various electrodynamic processes \cite{4, 5, 13, 16,17}. In the
field theory standard procedure of calculation of any value
representing interest from the  physical point of view, is based on
Feynman rules. The effective application of these rules
substantially simplifies calculations. The most important in Feynman
rules are expressions for propagator and tops describing structure
of the theory. We especially would like to emphasize that any
reorganization of the base of field theory first of all is a call to
quantum electrodynamics (QED)  single its   predictions excellently
be can coordinated to a number of high precisions experiments.
Therefore the hypothesis about FM at the beginning should be
considered with reference to QED. The requirement of gauge
invariance of QED with FM results in appearing of new interactions
which intensity is defined by the parameter of FM. Characteristic
feature of these interactions is obvious $ P$ and $CP$- symmetry
violation, and also not preservation of helicity at high-energy. Our
calculations show, that the basic test revealing distinction between
standard and new QED processes with the polarized particles.
Therefore we shall consider processes in view of particles
polarization. We have shown some experimental consequences of the
hypothesis about FM at high-energy \cite{4,5,16,17}. In comparison
with usual QED new Lagrangian contains additional members, which in
diagram technique should be with compared new vertices

$$ e(q+p)_{\mu}\gamma^{5} \quad and  \quad \frac{-\alpha}{\pi}
\gamma^{5}\sigma_{\mu \nu}.
$$

On basis of new Lagrangian QED, containing FM  M, we investigate the
section and asymmetry of  $ e^{-}e^{-} \to e^{-}e^{-}$ , $
e^{-}e^{+} \to e^{-}e^{+}$, $ e^{-}e^{+} \to \mu^{-} \mu^{+} $
processes, in view of particles polarization. It is received
differential sections
$$
\left[\frac{d\sigma}{d\Omega}\right]^{ e^{-}e^{-} \to e^{-}e^{-}}_{
\lambda_{1} \lambda_{2} \to
\lambda^{\prime}_{1}\lambda^{\prime}_{2}} \quad, \quad
\left[\frac{d\sigma}{d\Omega}\right]^{ e^{-}e^{+} \to e^{-}e^{+}}_{
\lambda_{1} \lambda_{2} \to
\lambda^{\prime}_{1}\lambda^{\prime}_{2}} and
 \left[\frac{d\sigma}{d\Omega}\right]^{ e^{-}e^{+} \to \mu^{-}\mu^{+}}_{ \lambda_{1} \lambda_{2} \to \lambda^{\prime}_{1}\lambda^{\prime}_{2}}
$$
which  contain  differential sections calculated on the basis of
standard QED and new members dependent on FM, where $\lambda_{1} $
and $\lambda_{2}$  are initial polarizations, $\lambda^{\prime}_{1}$
and $\lambda^{\prime}_{2}$ - final polarizations. For instance, the
asymmetry combination for the process  $ e^{-}e^{+} \to e^{-}e^{+}$
:
\begin{equation}
A=\frac{(sin^8 \frac{\theta}{2}+cos^8 \frac{\theta}{2})
(\frac{d\sigma}{d\Omega})_{\lambda_1=\lambda_2}-
(\frac{d\sigma}{d\Omega})_{\lambda_1=-\lambda_2} }
{(sin^8\frac{\theta}{2}+cos^8\frac{\theta}{2})
(\frac{d\sigma}{d\Omega})_{\lambda_1=\lambda_2}
+(\frac{d\sigma}{d\Omega})_{\lambda_1=-\lambda_2}}
\end{equation}
in the usual QED is different from zero due to the radiation
corrections but decreases at large
$$
E^{2} \, \approx \,\frac{\alpha m^2}{E^4}ln\frac{E^2}{m^2}.
$$
At the energies of the large electron-positron storage ring
accelerator at CERN, the radiation corrections will be negligibly
small. Then the main contribution to the quantity A will come from
the new interaction provided the FM varies from $M=1000GeV$. If
helicities of beginning particles are identical
$\lambda_{1}=\lambda_{2}$, process of annihilation $ e^{-}e^{+} \to
\mu^{-} \mu^{+} $ in usual QED follows the account of radiating
corrections and cross section decreases with energy of colliding
particles. However in our case we have
$$\left( \sigma _{tot} \right) ^{e^{+} e^{-}
\rightarrow \mu ^{+} \mu ^{-} } =\frac{\pi }{3} \frac{\sigma }{M^{2}
}.
$$

 On the basis of new QFT containing of FM cross  sections
and asymmetry of processes  $ e^{-}e^{-} \to e^{-}e^{-}$ , $
e^{-}e^{+} \to e^{-}e^{+}$ and $ e^{-}e^{+} \to \mu^{-} \mu^{+} $
are investigated in view of polarization of particles. Investigation
of differential cross sections and also asymmetry shows that in the
new circuit there are the effects caused by existence of FM. New
interaction does not keep spirality. In the theory $P$ and
$CP$-symmetry violation caused by existence of electric dipole
moments (EDM) of charged particles in particular: $d={e\ell/2}$.
Therefore experimental detection of EDM at electron, $\mu$-mesons
and $\tau$- lepton would have extremely important value for the
developed approach.

For more detailed research QFT with FM and revealings of
contribution FM at the first level it is necessary to investigate
cross sections of processes of Compton's scattering
$\gamma+e^{-}=\gamma^{ \prime}+e^{- \prime}$, annihilation process
in photon $e^{-}+e{+}=\gamma+\gamma$ and deep-inelastic dispersion
electron of on a nucleon. If to take into account the  on
accelerators of the future generation, to receive the polarized
particles in these processes, we shall take into account
polarization of particles.

As we have noted, any reorganization of the base of field theory
first of all is a call to the predictions of QED. Therefore in the
next stage of the investigation we shall apply to the hypothesis
about FM mass with reference to QED. To investigate consequences of
application of a hypothesis about FM in the following processes:
Compton's scattering $\gamma+e^{-}=\gamma^{ \prime}+e^{- \prime}$,
annihilation process of electron -positron pair in photon
$e^{-}+e^{+}=\gamma+\gamma$ and deep-inelastic dispersion of
electrons on a nucleon. These processes will be investigated taking
into account the polarizing effects in the second order of the
perturbation theory. For calculations we shall use usual engineering
of calculation for these processes. But thus we shall take into
account also new electrodynamics vertices in rules and Feynman
diagrams. Therefore in matrix elements of these processes new
members of the diagram will appear. In the received sections we
shall allocate a part dependent on FM. Results of our calculations
will be  compare with the  calculations at values $M \rightarrow
\infty $ according to standard QED. Thus will come to light a new
effects dependent on FM. If in a nature there will be this parameter
- FM, then a prediction of the given theory in the future by
experimenters  will be proved or justified.

{\bf Acknowledgments}\\
Author would like to thank V.G. Kadyshevsky, V. Gogokhia, P. Levai
and B. Lukacs, for valuable discussions, remarks and for their warm
hospitality. This work was supported by the MTA-JINR Collaboration
Grant.


\begin{thebibliography}{}

\bibitem{1}
Watagin G.V., Zs. Phez. 88, (1934), p.92 ; Heisenberg W., Zs. Phys.
101 (1936), p.533;  Heisenberg W., {\em Introduction to the Unifed
Field Theory of Elementtary Partyclies}, (Iters. Publ., (1966));
Snyder H., Phys. Rev. 71 (1947), p.38;{\em ibid} 72, 1947, p.68;
Yang C.N. Phys. Rev. 72 (1947), p.874; Markov M.A., Nucl. Phys. 10
(1958), p.140; Komar A.A. and Markov M. A. Nucl. Phys. 12 (1959),
p.190; Kadyshevsky V. G JETF 41 (1961), p.1885; DAN SSSR 147 (1962),
p.558;  Tamm I. E., Proceedings of XII Intern. Conference on High
Energy Physics,  Atomizdat, Moscow v.II, 1964 p.229; Mir-Kasimov R.
M., JETF 49 (1965), p.905;  Alebastrov V. A. and Efimov G. V., Comm.
jf Math. Physics 31,(1973), p.1; Bjorken J. D., {\em Indentification
of Elementary Forces and Gauge Theories}, (Harwood Academic
Publishere, (1979) p.701);  Donkov A. D., Kadyshevsky, V. G., Mateev
M. D. and Mir-Kasimov R. M., Bulgar. Journ. of Physics 1 (1974),
p.58, p.150, p.233; {\em ibid} 2 (1975), p3; Volobuyev I. P.,
Mir-Kasimov R. M., Acta Physica Polinica B9 (1978) p.2; Kadyshevsky
V.G., Nuclear Physics, B141 (1978), p.477; Kadyshevsky V.G.,
Particles and Nuclei, II, i.1 (1980), p.5;  Kadyshevsky V.G., Mateev
M.D., Phys. Lett. 106B (1980), p.5.
\bibitem{2}
Kadyshevsky V.G., Mateev M.D., {\em Quantum Field Theory and a New
Universal High Energy Scale. Scalar Fields}, Nuovo Cimento. 87A
(1985), p.324.
\bibitem{3}
Donkov A.D., Ibadov R.M., Kadyshevsky V.G., Mateev M.D. and Chizhov
M.V., {\em Some experimental consequences of a hypothesis about
fundamental length}, Izvestiya AN USSR, Series Physics T.46, N 9
(1982), p.1772.
\bibitem{4}
Ibadov R.M., Chizhov M.V., {\em Application of quantum
electrodynamics with In the fundamental length to high power
process},  Izvestiya AN UZSR, Series Physics and Mathematics,  N 5
(1983), p.38.
\bibitem{5}
Ibadov R.M., {\em Deep elastic scattering of Electron in Nucleons
and Fundamental lengths},  Izvestiya AN UZSR, Series Physics and
Mathematics, N 3 (1984), p.44.
\bibitem{6}
Chizhov M.V., Donkov A.D., Ibadov R.M., Kadyshevsky V.G. and Mateev
M.D., {\em Quantum Field Theory and a New Universal High Energy
Scale.  Dirac Fields}, Nuovo Cimento.  87A, No.3 (1985), p.350.
\bibitem{7}
Chizhov M.V., Donkov A.D., Ibadov R.M., Kadyshevsky V.G. and Mateev
M.D., {\em Quantum Field Theory and a New Universal High Energy
Scale. Gauge Vector Fields}, Nuovo Cimento.  V.87A, No.4  (1985),
p.375.
\bibitem{8}
Ibadov R.M., Kadyshevsky V.G, {\em About transformations of
supersymmetry in Theories of a Field with Fundamental Mass},
Preprint JINR. 2-86-835 Dubna (1986), 4 p.
\bibitem{9}
Ibadov R.M., Kadyshevsky V.G., {\em The new point of view on
auxiliary fields in supersymmetric models}, Works of VIII
International Meeting on problems of the Quantum Theory of a Field.
Alushta.  Dubna, D2-87-798, Dubna (1988), p.141.
\bibitem{10}
Ibadov R.M., Kadyshevsky V.G., {\em New formulation of Quantum field
theory with Fundamental mass}, in: Proceedings 5th International
Symposium on Selected Topics in Statistical Mechanics, Dubna, (World
Scientific, Singapore  (1989), p.131).
\bibitem{11}
Gel'fand I.M., Graev M.I., Vilenkin P.Ya., {\em Integral Geometry
and Related Problems of the Theory of Representations, Generalised
Functions}, i. 5, Fizmatgiz, M. (1962).
\bibitem{12}
Fursaev D.V., Kadyshevsky V.G., Ibadov R.M., {\em The Rotation
Invariant gauge Model with the compact momentum space}, in:
Proceedings of the 25th International Conference on High Energy
Physics, World Scientific, Singapore  (1990), p.928.
\bibitem{13}
Ibadov R.M., Ibadova U.R., Ravshanova D.R., Umidullaev Sh.U., {\em
On Quantum Field Theory with Fundamental mass}, in: Proc. of the
International Scholl-Seminar `` Structure of Particles and Nuclei
and Their Interactions `` . Humson, Dubna (1998), p.139.
\bibitem{14}
Umida Ibadova, Rustam Ibadov and Rashid Artikov, {\em Spontaneous
breaking of symmetry and fundamental mass},\\
http:pheno.physics.wisc.edu/pheno01 /ibadova.pdf
\bibitem{15}
Rustam M.Ibadov, Umida R.Ibadova, and Dilbar R.Ravshanova, {\em
Concept of Stochastic Quantization of the Abelian Fields and
Fundamental Mass},\\
http:pheno.physics.wisc.edu/pheno01/ravshanova.pdf.
\bibitem{16}
R.M.Ibadov, {\em New Quantum Field Theory}, Journal of Scientific
News, Samarkand, No.3, 2002, p.50.
\bibitem{17}
Ibadov R.M., {\em Hypothesis of Fundamental Mass and Concrete
Experimental Consequences at Ultrahigh Energies},in: Proceedings of
the II Eurasian Conference on Nuclear Science and ITS Application,
Republic of Kazakhstan, Almaty (2003), p.277.
\bibitem{18}
Ibadov R.M., {\em The Ratation Invariant Gauge Model with The
Compact Momentum Space},in: Proceeding of the 4-International
Conference Nuclear and Radition Physics, Kazakhstan, Almaty, 2003,
p.117.
\bibitem{19}
 Barnafoldi G.G., Leval P., Lukacs B.,
{\em Heavy quarks or compactified extra dimensions in the core of
hybrid stars,} arXiv:astro-ph/0312330 v1 12 Dec 2003;
 Lukacs  Bela,
Barnafoldi G. Gergely and Levai Peter, {\em The inner structure of
hybrid stars} arXiv:astro-ph/0312332 v1 12 Dec 2003.

\end{thebibliography}
\end{document}